\documentstyle[11pt]{article}   

\begin{document}
\setlength{\baselineskip}{4ex}  
 
\def\ds{\displaystyle}
\def\xp{x^\prime}
\def\tp{t^\prime}
\def\di{\partial}
\def\ni{\noindent}
\def\phib{\overline \phi}
\def\psib{\overline \psi}
\def\overphi{\overline \phi}
\def\ubar{{\overline u}}
\def\vbar{{\overline v}}
\def\overpsi{\overline \psi}
\def\overgam{\overline \gamma}
\def\overGam{\overline \Gamma}
\def\call{\cal l}
\def\calh{\cal h}
\def\dotphi{\dot\phi}
\def\dotchi{\dot\chi}
\def\bx{{\bf x}}
\def\bk{{\bf k}}
\def\br{{\bf r}}
\def\bp{{\bf p}}
\def\bq{{\bf q}}
\def\bra{{\langle}}
\def\ket{{\rangle}}

\def\vsl{{\vskip .6truecm}}
\def\vss{\vskip .4truecm}

\def\ds{\displaystyle}
\def\xp{x^\prime}
\def\rb{{\bf r}}
\def\pb{{\bf p}}
\def\pp{{p^{\prime}}}
\def\pbp{{{\bf p}^{\prime}}}
\def\bra{{\langle}}
\def\ket{{\rangle}}
\def\ep{{\varepsilon}}
\def\tp{t^\prime}
\def\di{\partial}
\def\ni{\noindent}
\def\phib{\overline \phi}
\def\psib{\overline \psi}
\def\oneha{{1\over 2}}
\def \vsl{\vskip .8cm}
\def\vss{\vskip .5cm}
\def\zo{{\overline Z}_1}
\def\zt{{\overline Z}_2}
\newcommand{\h}{\hat{H}}

\newcommand{\p}{{\bf{p}}}
\newcommand{\q}{{\bf{q}}} 
\newcommand{\k}{{\bf{k}}}

\def\ie{{\it i.e.~}}

\thispagestyle{empty}
\begin{center}

RELATIVISTIC THREE-FERMION WAVE EQUATIONS IN REFORMULATED QED 
AND RELATIVISTIC EFFECTS IN MUONIUM MINUS 

\end{center} 
\vskip 0.3truecm
\centerline{\large Mark Barham 
and Jurij W. Darewych  
}
\vskip 0.1truecm 
\centerline{\footnotesize\emph{Department of Physics and Astronomy, 
York University, Toronto, Ontario, M3J 1P3, Canada}} 
\vskip 0.2truecm 
\centerline{18 June 2007} 
\vskip 0.2truecm
\centerline{\textbf{Abstract}}

The variational method, within the Hamiltonian formalism of reformulated QED is used to 
determine relativistic wave equations for a system of three fermions of arbitrary 
mass interacting electromagnetically. 
The interaction kernels of the equations are, in essence, the invariant 
$\mathcal{M\,}$ matrices in lowest order. 
The equations are used to obtain relativistic $O(\alpha^2)$ corrections to the 
non-relativistic ground state 
energy levels of the Muonium negative ion ($\mu^+ e^- e^-$) as well as 
of $\mathrm{Ps}^-$ and $\mathrm{H}^-$,  using approximate variational three-body 
wave functions.  The results are 
compared with other calculations, where available. The relativistic correction 
for Mu$^-$  is found to be $-1.0773 \times 10^{-4}$ eV.


\section{Introduction}  
The bound state three-Fermion system, particularly Ps$^-$, has been the subject of 
theoretical investigations since the pioneering calculations of Wheeler \cite{wheeler} and 
Hylleraas \cite{hylleraas}, 
who first showed that this system has a single bound state. 
Although experimental measurements of the binding energy of Ps$^-$ have not been reported to date, 
there are preparations to make such measurements \cite{SF}.
\par 
Recently, Drake and Grigorescu 
reported an essentially exact (converged) variational calculation of the non-relativistic 
ground state energy of Ps$^-$ \cite{drake05}. They also used their very accurate 
wave function to calculate relativistic and QED 
corrections to the bound-state energy of this system. Accurate non-relativistic calculations 
of the Muonium negative ion (Mu$^-$ : $\mu^+ e^- e^-$) have been reported recently by Frolov \cite{frolov1}. 
Frolov used these to calculate the lowest-order QED $O(\alpha^3)$ corrections 
to the non-relativistic Mu$^-$ energy. However, relativistic ($O(\alpha^2)$)
corrections to the non-relativistic ground-state energy of Mu$^-$ seem not to 
have been calculated.
\par
In the present work, we work out a relativistic wave equation for a system of three fermions 
of arbitrary mass with electromagnetic interactions. This equation is used to obtain relativistic 
corrections to the bound-state energy of Mu$^-$, as well as of Ps$^-$ in order to compare 
our results to those of others.
\par

It has been shown in earlier works that a reformulation of various models in Quantum Field Theory (QFT), 
including QED, allows one to use simple Fock-state trial states to derive relativistic few-body wave 
equations by means of the variational method in the Hamiltonian formulation of the theory. 
An overview of this approach and various results obtained 
in this way for bosonic and fermionic systems (including Ps and Mu)
 is given in reference~\cite{Dar06} and citations therein. 
One of the advantages of this approach is that it permits straightforward generalization to 
relativistic systems of more than two particles.

\par


\section{Reformulated Hamiltonian formalism, field operators and variational method }  

The reformulated QED Hamiltonian density is \cite{LdeB, Dar06}
\begin{eqnarray}
{\cal H}_R &=& \sum_{a=1}^{3}\left[\psib_a(x)\left( -i\sum_{j=1}^3\gamma^j\frac{\di}{\di x^j} 
+m_a\right)\psi_a(x)-Q_a\psib_a(x)\gamma_\mu A^\mu_0(x)
\psi_a(x)\right]\nonumber\\
&+& {1\over 2}\int\! d^4\!\xp j^\mu(\xp) D_{\mu \nu}(x-\xp)
j^\nu(x),\label{1}  
\end{eqnarray}
where $\psi_a(x)$ are Dirac fermion fields of mass $m_a$ and charge $Q_a$, $A_0^\mu$ are free photon fields, 
\begin{equation}
j^\nu (x) = -\sum_{a=1}^{3}Q_a\;\overline \psi_a(x) \gamma^\nu \psi_a(x)
\label{eq2.4}
\end{equation}
are the fermionic source currents, and
$D_{\mu \nu}(x-\xp)= D_{\nu \mu}(x-\xp) = D_{\mu \nu}(\xp-x)$ are symmetric Green functions (photon propagators) defined by
\begin{equation}
\di_\alpha \di^\alpha D_{\mu \nu} (x-\xp) - \di_\mu \di^\alpha 
D_{\alpha \nu} (x - x^\prime )=g_{\mu \nu} \delta^4(x-\xp).
\label{eq2.6}
\end{equation}
In practice, one needs to choose a gauge, however, we do not need to specify one at this point.
\par
The reformulated Hamiltonian (\ref{1}) is obtained from the usual Lagrangian of QED by 
using the equations of motion to express the mediating photon field in terms of the 
fermion fields and photon field Green functions \cite{Dar06}, \cite{LdeB}. The reason 
for using the reformulated Hamiltonian is that it allows one to derive relativistic few-fermion wave 
 equations with the simplest possible Fock-space trial states.
Our notation is
\begin{equation} 
\psi_a(x) = \sum_{s=1}^2\int d^3p \,\frac{1}{(2\,\pi)^{{3\over
2}}}\sqrt{\frac{m_a}{\omega_{ap}}}\left[b_a(\bp , s) u_a(\bp , s) e^{-i p
\cdot x}+d_a^{\dag} (\bp , s) v_a(\bp , s) e^{i p \cdot x}\right], 
\label{eq2.13}
\end{equation}
where $p^\nu = \left(\omega_{ap} = \sqrt {m_a^2 + \bp^2},\bp\right)$. The mass-$m_a$ 
free-particle Dirac spinors $u_a$ and $v_a$, where $({\not\! p} -m_a) u_a(\bp,s)=0$ and  
$({\not\! p} +m_a) v_a(\bp,s)=0$, satisfy the following orthogonality conditions:
\begin{eqnarray}
u_a^{\dag}(\bp, s) u_a(\bp, \sigma) &=& v_a^{\dag}(\bp, s)
v_a(\bp,\sigma)=\frac{\omega_{ap}}{m_a}
\delta_{s\,\sigma}\label{eq2.14a}\\
u_a^{\dag}(\bp, s) v_a(-\bp, \sigma) &=& v_a^{\dag}(\bp, s)
u_a(-\bp, \sigma) = 0 .\label{eq2.14b}
\end{eqnarray}
The operators $b_a^{\dag}$ and $b_a$ are the creation and annihilation
operators for free particles of mass $m_a$; likewise, $d_a^{\dag}$ and
$d_a$ are the corresponding operators for antiparticles of mass $m_a$.  
These operators satisfy the usual anticommutation relations. The
non-vanishing ones are
\begin{equation}
\left\{b_a(\bp, s), b_a^{\dag}(\bq,\sigma)\right\} = 
\left\{d_a(\bp, s), d_a^{\dag}(\bq,\sigma)\right\} = 
\delta_{s \, \sigma}\delta^3 (\bp - \bq).\label{eq2.15}
\end{equation}  
As usual, operators for a given field commute with all the operators
corresponding to other fields.
\par
We use the above definitions to express the Hamiltonian operator, 
${\hat H_R} = \int d^3x \, {\cal H}_R$, 
in terms of the fermionic creation and annihilation operators, and we normal order the entire 
Hamiltonian (thereby denoting it by $:{\hat H}:$) in order to circumvent the need for vacuum 
and mass renormalization. We do not exhibit the Fourier decomposition of the photon field, 
since this is not needed in the present work.
\par
Since exact eigenstates of the Hamiltonian ${\hat H}_R $ \big(c.f.
 eq. (\ref{1})\big) are not obtainable, 
we determine approximations using the variational principle
\begin{equation} 
\delta \langle \Psi_{\rm tr} | :{\hat H}: - M | \Psi_{\rm tr} \rangle_{t=0} = 0.\label{22}
\end{equation}
%

%
%

\section{Relativistic three-fermion wave equations} 

For  systems of three fermions
 we use the following simple Fock-space trial state,
\begin{equation}   
|\Psi_{\rm tr} \ket = \sum_{{s_1}\,{s_2}\,{s_3}}\int d^3p_1 d^3p_2d^3p_3 \, 
F_{{s_1}\,{s_2}\,{s_3}}(\bp_1,\bp_2,\bp_3)\; b_1^{\dag}(\bp_1, s_1)b_j^{\dag}
(\bp_2,s_2)d_k^{\dag}(\bp_3, s_3) |0\ket ,       \label{eq3.1}
\end{equation}
where $F_{s_1 s_2 s_3}(\bp_1,\bp_2,\bp_3)$ are eight adjustable functions, and the vacuum state 
$|0\ket$ is defined by $b_j |0\ket = d_j |0\ket = 0$ for $j=1, 2, 3$.
We consider three cases for the values of $j$ and $k$: $j=1$
and $k=1$, $j=1$ and $k=3$, or $j=2$ and $k=3$. 
In the first case, the system consists of three particles of identical mass ({\sl e.g.}  $e^-e^-e^+$). 
In the second case, the system consists of two identical particles and a different antiparticle 
({\sl e.g.}  $e^-e^-\mu^+$). In the third case, the system consists of three distinct particles 
({\sl e.g.} $e^-\tau^-\mu^+$). 

\par
Substituting the trial state (\ref{eq3.1}) into (\ref{22})
we obtain the following relativistic momentum-space wave 
equations for the states of the three-fermion system:
\begin{eqnarray}                  
(\omega_{1q_1}&+&\omega_{jq_2}+ \, \omega_{kq_3}-E ) \,F_{r_1 r_2r_3}
(\bq_1,\bq_2,\bq_3)\label{eq3.9}\\ 
&+& \frac{i}{2(2\pi)^3}\sum_{s_1 s_2}\int d^3p_1d^3p_2 \, F_{s_1 s_2 r_3}
(\bp_1,\bp_2,\bq_3)  \,   
 \delta^3(\bp_1-\bq_1-\bq_2+\bp_2) \, {\cal M}^1_{r_1 r_2 s_1 s_2}
(\bp_1,\bp_2,\bq_1,\bq_2)\nonumber\\
&-&\frac{i}{2(2\pi)^3}\sum_{s_1 s_2}\int d^3p_1d^3p_2 \, F_{s_1 r_2 s_2}
(\bp_1,\bq_2,\bp_2)  \,  
\delta^3(\bp_1-\bq_1+\bp_2-\bq_3) \, {\cal M}^2_{r_1 r_3 s_1 s_2}
(\bp_1,\bp_2,\bq_1,\bq_3)\nonumber\\
&-&\frac{i}{2(2\pi)^3}\sum_{s_1 s_2}\int d^3p_1d^3p_2F_{r_1 s_1 s_2}
(\bq_1,\bp_1,\bp_2)  \,   
 \delta^3(\bp_1-\bq_2+\bp_2-\bq_3) \, {\cal M}^3_{r_2 r_3 s_1 s_2}
(\bp_1,\bp_2,\bq_2,\bq_3)\nonumber\\
&-&\frac{i\delta_{j1}\delta_{k1}}{(2\pi)^3}\sum_{s_1 s_2}\int d^3p_1d^3p_2 \, F_{r_1 s_2 s_1}
(\bq_1,\bp_2,\bp_1)   \,   
 \delta^3(\bp_1-\bq_2+\bp_2-\bq_3) \, {\cal M}^4_{r_2 r_3 s_1 s_2}
(\bp_1,\bp_2,\bq_2,\bq_3)\nonumber\\
&=& 0\nonumber
\end{eqnarray}
where
\begin{eqnarray} 
{\cal M}^1_{r_1 r_2 s_1 s_2}(\bp_1,\bp_2,\bq_1,\bq_2)& = & -iQ_1Q_j\ubar_1 
(\bq_1,r_1)\gamma^\mu u_1(\bp_1,s_1)\ubar_j (\bq_2,r_2)\gamma^\nu u_j(\bp_2,s_2)
 \label{eq3.10}  \\ 
&\times&[D_{\mu\nu}(\omega_{1p_1}-\omega_{1q_1},\bp_1
-\bq_1)+D_{\mu\nu}(\omega_{jp_2}-\omega_{jq_2},\bp_2-\bq_2)],  \nonumber 
\end{eqnarray}
\begin{eqnarray} 
{\cal M}^2_{r_1 r_3 s_1 s_2}(\bp_1,\bp_2,\bq_1,\bq_3)& = & -iQ_1Q_k\ubar_1 
(\bq_1,r_1)\gamma^\mu u_1(\bp_1,s_1)\vbar_k (\bp_2,s_2)\gamma^\nu v_k(\bq_3,r_3)    
\label{eq3.11}  \\  
&\times&[D_{\mu\nu}(\omega_{1p_1}-\omega_{1q_1},\bp_1-\bq_1)+D_{\mu\nu}
(\omega_{kp_2}-\omega_{kq_3},\bp_2-\bq_3)],    \nonumber
\end{eqnarray}
\begin{eqnarray}  
{\cal M}^3_{r_2 r_3 s_1 s_2}(\bp_1,\bp_2,\bq_2,\bq_3)& = & -iQ_jQ_k\ubar_j 
(\bq_2,r_2)\gamma^\mu u_j(\bp_1,s_1)\vbar_k(\bp_2,s_2)\gamma^\nu v_k(\bq_3,r_3)
\label{eq3.12}  \\  
&\times& [D_{\mu\nu}(\omega_{jp_1}-\omega_{jq_2},\bp_1-\bq_2)+D_{\mu\nu}(
\omega_{kp_2}-\omega_{kq_3},\bp_2-\bq_3)]   \nonumber  
\end{eqnarray}
are matrix elements corresponding to one-photon exchange Feynman diagrams in the 
particle-particle interaction, and for systems containing particle-antiparticle pairs (e.g.\ $e^+e^-e^-$)
\begin{eqnarray}  
{\cal M}^4_{r_2 r_3 s_1 s_2}(\bp_1,\bp_2,\bq_2,\bq_3)& = & iQ_1^2\ubar_1 
(\bq_2,r_2)\gamma^\mu v_1(\bq_3,r_3)\vbar_1(\bp_1,s_1)\gamma^\nu u_1(\bp_2,s_2)
\label{eq3.12a}  \\ 
&\times&
[D_{\mu\nu}(\omega_{1p_1}+\omega_{1p_2},\bp_1+\bp_2)
+D_{\mu\nu}(-\omega_{1q_2}-\omega_{1q_3},-\bq_2-\bq_3)]  \nonumber
\end{eqnarray}
is a matrix element corresponding to Feynman diagrams depicting virtual annihilation.
The virtual annihilation 
matrix elements are obtained along with one-photon exchange terms
  in the derivation and are not put in ``by hand''.
Higher order (loop) effects can be included by adding the appropriate ${\cal M}$-matrix elements 
to the kernels in Eq. (\ref{eq3.9}) or, more formally, by generalizing the trial state 
(\ref{eq3.1}), as was done for Ps \cite{TDH07}.

%
It is straightforward to verify that in the nonrelativistic limit, $(\bp /m_a)^2 \ll 1$,  
eq.~(\ref{eq3.9}) reduces to the usual three-body Schr\"odinger equation with Coulombic 
interparticle interactions. Details of this, as well as of all other calculations presented here, 
are given in reference \cite{BarhamThesis}.
\par
At this point it is worthwhile mentioning that the relativistic three-fermion 
eq.~(\ref{eq3.9}) holds for any values of the masses (\ie no recoil corrections 
are necessary) and any strength of the coupling. In addition, this equation, being 
Salpeter-like rather than Dirac-like, has only positive-energy solutions and is 
amenable to variational solution without any ``negative-energy'' difficulties.

It is impossible to solve eq.~(\ref{eq3.9}) analytically (even in the nonrelativistic limit).
Therefore, approximate (\ie numerical, variational or perturbative) solutions must be 
sought for various cases of interest. This is a 
non-trivial task even in the nonrelativistic case; hence all the more so for 
the relativistic eq.~(\ref{eq3.9}). 
We shall  set up the variational solution of eq.~(\ref{eq3.9}), however, in this paper, 
we will use the resulting matrix elements to calculate perturbatively the 
 (comparatively small) $O(\alpha^2)$ relativistic  
corrections to the non-relativistic energy eigenvalues for  
Mu$^-$, Ps$^-$ and H$^-$.


%
\section{Variational approximations and relativistic corrections to the bound-state energy 
of  Mu$^-$ and Ps$^-$}
 For variational approximations the trial state, eq.~(\ref{eq3.1}), can be chosen 
such that the eight adjustable functions 
take the following spin and momentum separable form
\begin{equation}
F_{{s_1}\,{s_2}\,{s_3}}(\bp_1,\bp_2,\bp_3)=\Lambda_{s_1s_2s_3}f(\bp_1,\bp_2,\bp_3),
\label{eq3.52} 
\end{equation} 
where $f(\bp_1,\bp_2,\bp_3)$ is an adjustable function and $\Lambda_{s_1s_2s_3}$ are a 
set of constants. For systems like Ps$^-$, Mu$^-$ of H$^-$ we consider the two cases,
\begin{enumerate}
\item $\Lambda_{111}=\Lambda_{221}=\Lambda_{s_1s_22}=0,\Lambda_{121}=
-\Lambda_{211}=1/\sqrt{2}$
for all $s_1$, $s_2$, $S=1/2, m_s=1/2$
\item $\Lambda_{112}=\Lambda_{222}=\Lambda_{s_1s_21}=0, \Lambda_{122}=
-\Lambda_{212}=1/\sqrt{2}$
for all $s_1$, $s_2$, $S=1/2, m_s=-1/2$
\end{enumerate}
where $S$ is the total spin and $m_s$ is the spin projection of the state.
For both cases, the spin part of the adjustable function is normalized
such that $\sum _{s_1s_2s_3}\Lambda^*_{s_1s_2s_3}\Lambda_{s_1s_2s_3}=1$. 
Thus, the trial state takes a form in which particles 1 and 2
are described by a spin singlet state; for case one particle 3 is in a 
spin up state and for case two particle 3 is in a spin down state. 
We consider the special cases where $j=1$, $k=1,2,3$, $Q_1=e$, $Q_k=Z_ne$
where $Z_n$ is a positive integer and $e$ is the elementary charge. The
cases with $Z_n=1$ correspond to systems like $e^-e^-e^+$, $e^-e^-\mu^+$
and $^1\mathrm{H}^-$. For the cases where $Z_n>1$, particle 3 may be thought
of as the nucleus of a Helium atom (i.e.\ $Z_n=2$) or a Helium-like ion 
(i.e.\ $Z_n>2$). For the cases in which the positively charged particle is 
the nucleus of an atom and not a fundamental fermion the results of the 
perturbative calculation will apply approximately to these systems if their 
total nuclear spin is 1/2, or if the nucleus is very massive and may be 
treated as a static charge (\ie the $m_3 \to \infty$ limit).  
\par
Multiplying eq.~(\ref{eq3.9}) by $F^*_{r_1 r_2 r_3}(\bq_1,\bq_2,\bq_3)$ and  
integrating over all $\bq_1,
\bq_2,\bq_3$, summing over all $r_1,r_2,r_3$ and applying the normalization 
condition $\sum_{s_1s_2s_3}\Lambda^*_{s_1s_2s_3}\Lambda_{s_1s_2s_3}=1$ we 
obtain the following expression for the energy,  
\begin{equation}
E=\bra {\hat H}_{0}\ket +\bra {\hat H}_{I12}\ket +2\bra {\hat H}_{I13}\ket ,
 \label{eq3.53}
\end{equation}
where ${\int d^3q_1d^3q_2d^3q_3\, f^*(\bq_1,\bq_2,\bq_3)f(\bq_1,\bq_2,\bq_3)}$ is 
taken to be unity (or, equivalently, the right-hand side of eq.~(\ref{eq3.53}) must be 
divided by this factor). The contributing matrix elements are
\begin{equation}
\bra {\hat H}_{0}\ket =\int d^3q_1d^3q_2d^3q_3\, f^*(\bq_1,\bq_2,\bq_3)
f(\bq_1,\bq_2,\bq_3)\Big[\omega_{1q_1}+\omega_{1q_2}+\omega_{3q_3}\Big],
\label{eq3.54}
\end{equation}
\begin{eqnarray}
\bra {\hat H}_{I12}\ket &=& \frac{e^2}{2(2\pi)^3}\int d^3p_1d^3p_2d^3q_1d^3q_2
d^3q_3f^*(\bq_1,\bq_2,\bq_3)f(\bp_1,\bp_2,\bq_3)\nonumber\\
&\times&\delta^3(\bp_1-\bq_1-\bq_2+\bp_2)K^{\mu\nu}_{12}(\bp_1,\bp_2,\bq_1,
\bq_2)\nonumber\\
&\times&[D_{\mu\nu}(\omega_{1p_1}-\omega_{1q_1},\bp_1-\bq_1)+D_{\mu\nu}(
\omega_{1p_2}-\omega_{1q_2},\bp_2-\bq_2)],\label{eq3.55}
\end{eqnarray}
\begin{eqnarray}
\bra {\hat H}_{I13}\ket &=& -\frac{Z_ne^2}{2(2\pi)^3}\int d^3p_1d^3p_2d^3q_1
d^3q_2d^3q_3 f^*(\bq_1,\bq_2,\bq_3)f(\bp_1,\bq_2,\bp_2)\nonumber\\
&\times&\delta^3(\bp_1-\bq_1+\bp_2-\bq_3)K^{\mu\nu}_{13}(\bp_1,\bp_2,\bq_1,
\bq_3)\nonumber\\
&\times&[D_{\mu\nu}(\omega_{1p_1}-\omega_{1q_1},\bp_1-\bq_1)+D_{\mu\nu}(
\omega_{3p_2}-\omega_{3q_3},\bp_2-\bq_3)],\label{eq3.56}
\end{eqnarray}
\begin{eqnarray}
K^{\mu\nu}_{12}(\bp_1,\bp_2,\bq_1,\bq_2) &=& B_{12}(\bp_1,\bp_2,\bq_1,\bq_2)
[K_1^\mu (\bq_1,\bp_1,m_1)K_1^\nu (\bq_2,\bp_2,m_1)\nonumber\\
&-& K_2^\mu (\bq_1,\bp_1,m_1)K_2^\nu (\bq_2,\bp_2,m_1)\nonumber\\
&-&K_3^\mu (\bq_1,\bp_1,m_1)K_3^\nu (\bq_2,\bp_2,m_1)\nonumber\\
&+& K_4^\mu (\bq_1,\bp_1,m_1)K_4^\nu (\bq_2,\bp_2,m_1)],\label{eq3.57}
\end{eqnarray}
\begin{eqnarray}
B_{12}(\bp_1,\bp_2,\bq_1,\bq_2) &=&
\frac{1}{4\sqrt{\omega_{1q_1}\omega_{1p_1}\omega_{1q_2}\omega_{1p_2}}}\nonumber\\
&\times& \frac{1}{\sqrt{(\omega_{1q_1}+m_1)(\omega_{1p_1}+m_1)(\omega_{1q_2}+m_1)
(\omega_{1p_2}+m_1)}},\label{eq3.58}
\end{eqnarray}
\begin{eqnarray}
K^{\mu\nu}_{13}(\bp_1,\bp_2,\bq_1,\bq_3)&=&B_{13}(\bp_1,\bp_2,\bq_1,\bq_3)
K_1^\mu(\bq_1,\bp_1,m_1)[K_1^\nu (\bp_2,\bq_3,m_3)\nonumber\\
&\mp& K_2^\nu (\bp_2,\bq_3,m_3)],\label{eq3.59}
\end{eqnarray}
\begin{eqnarray}
B_{13}(\bp_1,\bp_2,\bq_1,\bq_3) &=&
\frac{1}{4\sqrt{\omega_{1q_1}\omega_{1p_1}\omega_{3p_2}\omega_{3q_3}}}\nonumber\\
&\times& \frac{1}{\sqrt{(\omega_{1q_1}+m_1)(\omega_{1p_1}+m_1)(\omega_{3p_2}+m_3)
(\omega_{3q_3}+m_3)}},\label{eq3.60}
\end{eqnarray}
\begin{equation}
K_1^\mu (\bp ,\bq ,m_a)=g^{0\mu}(m_a^2-\omega_{ap}\omega_{aq}+\bp\cdot\bq
) +p^\mu (m_a+\omega_{aq})
+q^\mu (m_a +\omega_{ap}),\label{eq3.61}
\end{equation}
\begin{eqnarray}
K_2^\mu (\bp ,\bq ,m_a) &=& 
i(g^{0\mu}(\bp_1\bq_2-\bq_1\bp_2)+g^{1\mu}[\bp_2(m_a+\omega_{aq})-\bq_2
(m_a+\omega_{ap})]\nonumber\\
&+& g^{2\mu}[\bq_1(m_a+\omega_{ap})-\bp_1(m_a+\omega_{aq})]),\label{eq3.62}
\end{eqnarray}
\begin{eqnarray}
K_3^\mu (\bp ,\bq ,m_a) &=& 
i(g^{0\mu}(\bp_2\bq_3-\bq_2\bp_3)+g^{2\mu}[\bp_3(m_a+\omega_{aq})-\bq_3
(m_a+\omega_{ap})]\nonumber\\
&+& g^{3\mu}[\bq_2(m_a+\omega_{ap})-\bp_2(m_a+\omega_{aq})]),\label{eq3.63}
\end{eqnarray}
\begin{eqnarray}
K_4^\mu (\bp ,\bq ,m_a) 
&=& g^{0\mu}(\bp_1\bq_3-\bq_1\bp_3)+g^{1\mu}[\bp_3(m_a+\omega_{aq})-\bq_3
(m_a+\omega_{ap})]\nonumber\\
&+& g^{3\mu}[\bq_1(m_a+\omega_{ap})-\bp_1(m_a+\omega_{aq})],\label{eq3.64}
\end{eqnarray}
$i=\sqrt{-1}$, $a=1,2$, $p^0 =\omega_{ap}$ and $p^j=\bp_j$ where $j=1,2,3$. 
Note that the subscripts on the vectors in equations~(\ref{eq3.61})-(\ref{eq3.64}), 
unlike elsewhere, 
 denote the components of the generic vectors $\bp$ and $\bq$. 
\par
 The sign $\mp$ in eq.~(\ref{eq3.59}) are taken to be $-$ if particle 3 has 
spin projection $m_s=1/2$ (\ie \ spin up) or $+$ if particle 3 has spin projection 
$m_s=-1/2$ (\ie \ spin down). Also note that the matrix 
element corresponding to the interaction between particles 1 and 3 is identical to 
the matrix element corresponding to the interaction between particles 2 and 3 
(particles 1 and 2 are identical so that their respective 
interactions with particle 3 provide identical contributions to the energy); 
hence the factor $2$ in front of $\bra {\hat H}_{I13}\ket$ in eq.~(\ref{eq3.53}). 
\par
In practice, calculation are done in the rest-frame, for which
$f(\bp_1,\bp_2,\bp_3)=\delta^3(\bp_1+\bp_2+\bp_3)f(\bp_1,\bp_2)$ where $f(\bp_1,\bp_2)$  
is an adjustable function (normalized to unity). So far no assumptions 
about the adjustable function $f(\bq_1,\bq_2,\bq_3)$, or $f(\bp_1,\bp_2)$ 
in the rest frame, have 
been made. For relativistic variational approximations valid at arbitrary 
strength of the coupling,  $f$ would be expressed by analytic forms with adjustable 
features (parameters), which would be chosen to minimize the expectation 
value of the Hamiltonian (eq.~(\ref{eq3.53})). However, as already stated, 
we shall not pursue 
such a variational approach in this work. Instead, we shall obtain perturbative 
solutions which are valid for weak coupling.
%
\par
To obtain the order $\alpha^4$ contributions to the three-fermion energy we expand 
$\omega_{jp}$ and all kernels in the above equations to lowest order beyond 
their non-relativistic limit (the explicit forms are given in 
ref. \cite{BarhamThesis}). We use the Coulomb gauge. 
The resulting expression for the energy is 
\begin{equation}
E=2m_1+m_3+E_0+\Delta E\label{eq3.76}   
\end{equation}
where
\begin{eqnarray}
E_0 &=& \int d^3q_1d^3q_2\left[\frac{\bq_1^2}{2m_1}+\frac{\bq_2^2}{2m_1}+
\frac{|\bq_1+\bq_2|^2}{2m_3}\right]|f(\bq_1,\bq_2)|^2\nonumber\\
&-& \frac{2Z_ne^2}{(2\pi )^3}\int\!\!
d^3p_1d^3q_1d^3q_2\frac{f^*(\bq_1,\bq_2)f(\bp_1,\bq_2)}
{|\bp_1-\bq_1|^2}\nonumber\\
&+& \frac{e^2}{(2\pi )^3}\int\!\!
d^3p_1d^3q_1d^3q_2\frac{f^*(\bq_1,\bq_2)f(\bp_1,\bq_1+\bq_2-\bp_1)}
{|\bp_1-\bq_1|^2}\label{eq3.77}
\end{eqnarray}
and
\begin{equation}	
\Delta E=\Delta KE +\sum_{i=1}^3\Delta PE_{12i} +2 \sum_{i=1}^4\Delta PE_{13i},\label{eq30}
\end{equation}
where
\begin{equation}
\Delta KE = -\frac{1}{8}\int d^3q_1d^3q_2\left[\frac{\bq_1^4}{m_1^3}+\frac{\bq_2^4}{m_1^3}+
\frac{|\bq_1+\bq_2|^4}{m_3^3}\right]|f(\bq_1,\bq_2)|^2,\label{eq3.79}
\end{equation}
\begin{equation}
\Delta PE_{131}=\frac{Z_ne^2}{8(2\pi )^3}\left(\frac{1}{m_1^2}+\frac{1}{m_3^2}\right)
\int\!\! d^3p_1d^3q_1d^3q_2f^*(\bq_1,\bq_2)f(\bp_1,\bq_2),\label{eq3.82}
\end{equation}
\begin{equation}
\Delta PE_{132}=-\frac{Z_ne^2}{m_1m_3(2\pi )^3}
\int\!\! d^3p_1d^3q_1d^3q_2f^*(\bq_1,\bq_2)f(\bp_1,\bq_2)\frac{|\bp_1\times\bq_1|^2}
{|\bp_1-\bq_1|^4},\label{eq3.83}
\end{equation}
\begin{equation}
\Delta PE_{133}=-\frac{Z_ne^2}{2m_1m_3(2\pi )^3}
\int\!\! d^3p_1d^3q_1d^3q_2f^*(\bq_1,\bq_2)f(\bp_1,\bq_2)\frac{(\bp_1+\bq_1)\cdot\bq_2}
{|\bp_1-\bq_1|^2},\label{eq3.84}
\end{equation}
\begin{equation}
\Delta PE_{134}=\frac{Z_ne^2}{2m_1m_3(2\pi )^3}
\int\!\! d^3p_1d^3q_1d^3q_2f^*(\bq_1,\bq_2)f(\bp_1,\bq_2)\frac{(\bp_1^2-\bq_1^2)(\bp_1-\bq_1)
\cdot\bq_2}{|\bp_1-\bq_1|^4},\label{eq3.85}
\end{equation}
\begin{equation}
\Delta PE_{121}=\frac{e^2}{4m_1^2(2\pi )^3}
\int\!\! d^3p_1d^3q_1d^3q_2f^*(\bq_1,\bq_2)f(\bp_1,\bq_1+\bq_2-\bp_1),\label{eq3.86}
\end{equation}
\begin{equation}
\Delta PE_{122}=-\frac{e^2}{2m_1^2(2\pi )^3}
\int\!\! d^3p_1d^3q_1d^3q_2f^*(\bq_1,\bq_2)f(\bp_1,\bq_1+\bq_2-\bp_1)
\frac{(\bp_1+\bq_1)\cdot\bq_2}{|\bp_1-\bq_1|^2},\label{eq3.87}
\end{equation}
\begin{eqnarray}
\Delta PE_{123}&=&\frac{e^2}{2m_1^2(2\pi )^3}\int\!\! d^3p_1d^3q_1d^3q_2
f^*(\bq_1,\bq_2)f(\bp_1,\bq_1+\bq_2-\bp_1)   \nonumber     \\
&\times&
\frac{(\bp_1^2-\bq_1^2)(\bp_1-\bq_1)\cdot\bq_2}{|\bp_1-\bq_1|^4}.
\label{eq3.88}
\end{eqnarray}
Note that the expressions for the energy in equations~(\ref{eq3.76})-(\ref{eq3.88}) 
do not depend on the spin projection of particle 3; therefore, both trial states 
 yield the same kinetic, potential and total energy.

%
\par
In order to evaluate perturbatively the relativistic corrections, $\Delta E$, 
from equations~(\ref{eq3.79}), 
(\ref{eq3.82})-(\ref{eq3.88}),  
$f(\bp_1,\bp_2)$ should be a 
solution of the three-body Schr\"{o}dinger equation. 
However, exact solutions of
this equation are not available; therefore, we shall use simple variational wave functions
 that will 
allow for the approximate evaluation of the non relativistic expression $E_0$
for the energy in eq.~(\ref{eq3.77}) and the relativistic correction terms 
in equations~(\ref{eq3.79}), (\ref{eq3.82})-(\ref{eq3.88}). 
\par
The  $\mu^+ e^- e^-$, Ps$^-$ and H$^-$ ions have only one bound state, namely 
the ground state, which we shall represent by the simple (but sufficient for our purposes) 
wave function with two distance-scale parameters. In coordinate representation this wave 
function is
\begin{equation}
\psi_t (\bx_1,\bx_2)=\frac{1}{\sqrt{N}}[\phi_{100}(\bx_1,\zo )\phi_{100}(\bx_2,\zt )
+\phi_{100}(\bx_1,\zt )\phi_{100}(\bx_2,\zo )  ,   \label{eq4.2.18}
\end{equation}
where 
\begin{equation}
\phi_{100}(\bx_i,{\overline Z}_j)=R_{10}(x_i,{\overline Z}_j)
Y_0^0(\theta_i,\phi_i), ~~~~
R_{10}(x_i,{\overline Z}_j)=2\sqrt{a_j^3}e^{-a_jx_i},
\label{eq4.2.19}
\end{equation}
($i,j=1,2$), $a_j={\overline Z}_j\mu\alpha$ and $N$ is the normalization factor. 
The wave function, Eq.~(\ref{eq4.2.18}), consists of hydrogenic $1s$ forms 
for the two electrons but with two different distance scale parameters $\zo$ and $\zt$. 
The explicit expressions for $E_0$ and $\Delta E$ as functions of the parameters 
$\zo$ and $\zt$ are given in the Appendix.

%
\section{Numerical results and discussion}

The minimum value of $E_0(\zo,\zt)$ and corresponding values of the variational parameters 
for Mu$^-$, as well as for Ps$^-$ 
 and H$^-$ are given in Table 1. 
We use the values $m_e =510999.137$ eV and $\alpha =1/137.03599911$ 
and the conversion factors  1 au=27.2113962 eV and 1 Ry=13.6056981 eV. 
The values of the scale parameters, which are  
$\zo \simeq 1$ and $\zt \simeq 0.28$  for all three systems, indicate that 
each can be pictured as an electron orbiting a neutral atom.
%
%
We also list very accurate variational 
energies obtained with many parameter wave functions 
by Drake and Grigorescu~\cite{drake05} for Ps$^-$ and by Frolov~\cite{frolov1}
for Mu$^-$ and H$^-$. 
\par
Note that the simple variational predictions of the non-relativistic 
ground state energies differ from the very accurate values by 
2.04\%, 2.71\% and 2.74\% for Ps$^-$, Mu$^-$ and H$^-$ respectively. 
This implies that the relativistic $O(\alpha^4)$ corrections calculated with the simple 
wave function Eq.~(\ref{eq4.2.18}) will be uncertain by a corresponding amount.
%
%
\begin{table}
\caption{Non-relativistic variational energies $E_0$ and the corresponding 
optimum values for$\zo$, $\zt$, along with very accurate values obtained 
from the literature,  for the ground states of Ps$^-$, 
Mu$^-$ and $^1\mathrm{H}^-$. $E_0$ is expressed in eV and converted 
to atomic units (1 au=27.2113962 eV) or Rydbergs (1 Ry=13.6056981 eV). 
The terms in brackets for $^1\mathrm{H}^-$ are results 
obtained for the case where the mass of the nucleus is assumed to be infinite.}
\begin{center}
\begin{tabular}{|c|c|}\hline
$\mathrm{Ps}^-$ ground state & Value \\ \hline
$m_3$ (rest mass energy of a positron) & 510999.137 eV \\
$\zo$ & 1.03922997 \\  
$\zt$ & 0.283221430 \\
$E_0$ & -6.98384409 eV=-0.513302885 Ry \\ 
energy from Drake and Grigorescu~\cite{drake05} & -0.52401014046596021539 Ry \\ 
binding energy of $e^+e^-$ & -6.80284905 eV \\ \hline
$\mathrm{Mu}^-$ ground state & Value \\ \hline
$m_3$ (rest mass energy of $\mu^+$) & 105658403 eV \\
$\zo$ & 1.03922997 \\
$\zt$ & 0.283221432 \\
$E_0$ & -13.9004610 eV=-0.510832331 au \\
energy from Frolov~\cite{frolov1} & -0.5250548062435263292914 au \\ 
binding energy of $\mu^+e^-$ & -13.5402131 eV \\ \hline
$^1\mathrm{H}^-$ ground state & Value \\ \hline
$m_3$ (rest mass energy of $^1\mathrm{H}^-$ nucleus) & 938272446 eV \\
$\zo$ & 1.03922997 \\
 & (1.03922997) \\
$\zt$ & 0.283221432 \\
 & (0.283221431) \\
$E_0$ & -13.9600853 eV=-0.513023483 au \\ 
 & (-13.9676882 eV=-0.513302885 au) \\
energy from Frolov~\cite{frolov1} & -0.5274458811141788934109 au \\
binding energy of $^1\mathrm{H}$ & -13.5982922 eV  \\
\hline
\end{tabular}
\end{center}
\label{table:non_rel1a}
\end{table}
The values of the relativistic $O(\alpha^4)$ contributions to the energy,  
 $\Delta E (\zo,\zt)$ (cf. Eq.(\ref{eq30})) for the three systems,  Ps$^-$, Muonium$^-$ 
and H$^-$, evaluated by using the wave function (\ref{eq4.2.18}), 
with the $\zo, \zt$  values of Table 1, are exhibited in Table  2.      
For Ps$^-$, we also list the results obtained by Drake and 
Grigorescu~\cite{drake05},  Frolov~\cite{frolov2} and Bhatia and Drachman 
\cite{bhatia:drachman2}. 
%
%
\begin{table}
\caption{Non-relativistic values for $\zo$, $\zt$ and $E_0$ 
(cf. Table~\ref{table:non_rel1a}) 
and the corresponding $O(\alpha^4)$ energy corrections $\Delta E$ (in eV), 
obtained by using these $\zo$, $\zt$ values, 
for the ground state of  
$\mathrm{Ps}^-$, $\mathrm{Mu}^-$ and $^1\mathrm{H}^-$. 
Results for $\Delta E$ for $Ps^-$ obtained by other workers 
are included for comparison.
} 
\begin{center}
\begin{tabular}{|c|c|c|c|}\hline
 & \multicolumn{3}{c|}{System} \\ \cline{2-4}
Term  & $\mathrm{Ps}^-$ & $\mathrm{Mu}^-$ & $\mathrm{H}^-$ \\ \hline
$\zo$ & 1.03922997 & 1.03922997 & 1.03922997 \\
$\zt$ & $2.83221430\times 10^{-1}$ & $2.83221432\times 10^{-1}$ & $2.83221432\times 10^{-1}$ \\
$E_0$ & -6.98384409 & -13.9004610 & -13.9600853 \\ 
$\Delta KE$ & $-1.11369869\times 10^{-4}$ & $-8.48276777\times 10^{-4}$ & 
$-8.62924906\times 10^{-4}$ \\ 
$\Delta PE$ & $0.11061221\times 10^{-4}$ & $7.40544121\times 10^{-4}$ & 
$7.56110074\times 10^{-4}$ \\ 
$\Delta E^{\rm ~this~ work}$ & $-1.00308648\times 10^{-4}$ & $-1.07732656\times 10^{-4}$ & 
$-1.06814831\times 10^{-4}$ \\
$\Delta E^{\rm Drake, Grigorescu \, \cite{drake05}}
 $ & $-1.054006746\times 10^{-4}$ &           &  \\
$\Delta E^{\rm Frolov \, \cite{frolov2}}
 $ & $-0.914436125\times 10^{-4}$ &           &  \\
$\Delta E^{\rm Bhatia, Drachman \, \cite{bhatia:drachman2}}
 $ & $-0.91702290\times 10^{-4}$ &           &  \\
\hline
\end{tabular}
\end{center}
\label{table:rel1a}
\end{table}
Examining the results presented in Table 2, we note that 
 the $O(\alpha^4)$ corrections for each of Ps$^-$, Mu$^-$ and H$^-$, are 
smaller in magnitude by a factor of the order of $\alpha^2$ 
in comparison to the non-relativistic energies $E_0$, as is to be expected and 
as happens also in the two-fermion systems Ps, Mu ($\mu^+ e^-$) and H.
\par
We note that the entries in Table 2 for Mu$^-$ and H$^-$ 
are quite similar, as one might expect, 
since $m_e / m_\mu$ and $m_e / m_p$ are both much less than 1, so that 
recoil effects are small. 
 It is interesting to note, however, that $\Delta E$ is very similar for all 
three systems, Ps$^-$, Muonium$^-$ and H$^-$  
even though 
kinetic and potential energy contributions differ substantially between Ps$^-$ 
on the one hand, and Mu$^-$ and H$^-$ on the other.
\par 
Our results for $\Delta E$ for Ps$^-$
agree quite well with the corresponding results obtained by  Drake and 
Grigorescu~\cite{drake05},  Frolov~\cite{frolov2} and Bhatia and Drachman 
\cite{bhatia:drachman2}. This suggests that our results for $\Delta E$ for Mu$^-$ 
are of reasonable accuracy as well. As far as we know no previous calculations of 
$\Delta E$ for Mu$^-$ have been reported in the literature. 
\par
At this time experimental measurements of the Ps$^-$ and Mu$^-$ 
binding energy are not available, although plans to make such measurements for Ps$^-$ 
are being considered \cite{SF}. We expect that measurements for Mu$^-$ will also 
be forthcoming in the future. 
\vskip .5cm
The financial support of the Natural Sciences and Engineering Research Council of Canada 
for this research is gratefully acknowledged.


%
\par  
\ni {\bf{ APPENDIX } }
\par
\ni Expectation values for the non-relativistic ground state energy:
\par
\ni ($m_1=m_2=m,~ m_3=M,~ \mu = mM/(m+M)$) 
\begin{equation}           
E_0(\zo,\zt) = \bra \hat{K} \ket  
+\bra\hat{V}_{13}\ket
+\bra\hat{V}_{23}\ket+\bra\hat{V}_{12}\ket\label{eq4.2.9}
\end{equation}
\begin{eqnarray}   
\bra \hat{K} \ket    
&=&\frac{\mu\alpha^2}{N}\left(\frac{26\zo^5\zt^3+158\zo^4\zt^4+16\zo^6\zt^2
+16\zo^2\zt^6}{(\zt +\zo )^6}\right.\nonumber\\
&+&\left.\frac{26\zo^3\zt^5+\zo^8+6\zo^7\zt+\zt^8+6\zt^7\zo}{(\zt +\zo )^6}\right) ,\label{eq4.2.21}
\end{eqnarray}
\begin{equation}
\bra \hat{V}_{13} \ket = \bra \hat{V}_{23} \ket =
- \frac{1}{2} Z_n \, \mu \, \alpha^2 (\zo+\zt).
\label{43}
\end{equation}
\begin{equation}
\bra\hat{V}_{12}\ket =2\mu\alpha^2\zo\zt\frac{28\zo^2\zt^2+5\zo\zt^3+\zo^4+\zt^4+5\zo^3\zt}
{N(\zt+\zo )^5}.\label{eq44}
\end{equation}
\begin{equation}
N=2 \left[ 1 + \frac{64 \zo^3 \zt^3}{(\zo+\zt)^6}\right] = 
2\frac{15\zo^4\zt^2+15\zt^4\zo^2+\zt^6+84\zo^3\zt^3+\zo^6+6\zo^5\zt+6\zt^5\zo}{(\zt+\zo )^6}.
\label{eq4.2.20a}
\end{equation}
\par
The minimum values of $E_0(\zo,\zt)$ and the corresponding values of $\zo$ and $\zt$ are given in Table 1.
\par
Expectation values for the relativistic corrections:
\begin{eqnarray}
\Delta KE &=&-\frac{\mu^4\alpha^4}{4N(\zt +\zo )^6}\left(\frac{1}{m^3}+\frac{1}{M^3}\right) (5\zo^{10}+30\zt\zo^9+75\zo^8\zt^2\nonumber\\ 
&+& 100\zt^3\zo^7
+208\zo^6\zt^4+444\zt^5\zo^5+208\zo^4\zt^6+100\zt^7\zo^3\nonumber\\
&+& 75\zo^2\zt^8+30\zt^9\zo+5\zt^{10})-\frac{5\mu^4\alpha^4\zo^2\zt^2}{6M^3N(\zt+\zo )^6}
(15\zo^4\zt^2\nonumber\\
&+& 15\zt^4\zo^2+\zt^6+84\zo^3\zt^3+\zo^6+6\zo^5\zt+6\zt^5\zo ),\label{eq.kegs}
\end{eqnarray}
%
\begin{eqnarray}
\Delta PE_{131} &=& \frac{\mu^3\alpha^4Z_n(M^2+m^2)}{2M^2m^2N(\zt +\zo )^3}
(\zo^6+3\zo^5\zt+3\zo^4\zt^2\nonumber\\
&+& 18\zo^3\zt^3+3\zt^4\zo^2+3\zt^5\zo+\zt^6),\label{eq.pe131gs}
\end{eqnarray}
\begin{eqnarray}
\Delta PE_{132} &=& -\frac{Z_n\mu^3\alpha^4}{mMN(\zt +\zo )^5}
(5\zt^7\zo+10\zo^6\zt^2+\zt^8+11\zo^5\zt^3\nonumber\\
&+& 5\zo^7\zt+74\zo^4\zt^4+11\zo^3\zt^5+10\zo^2\zt^6+\zo^8),\label{eq.pe132gs}
\end{eqnarray}
\begin{equation}
\Delta PE_{133}=\Delta PE_{134}=0,\label{eq.pe133and134gs}
\end{equation}
\begin{equation}
\Delta PE_{121}=\frac{4\mu^3\alpha^4\zo^3\zt^3}{m^2N(\zt +\zo )^3},\label{eq.pe121gs}
\end{equation}
\begin{equation}
\Delta PE_{122}=- \Delta PE_{123}= -\frac{2\mu^3\alpha^4\zo^3\zt^3(\zt -\zo )^2}{m^2N(\zt+\zo )^5},
\label{eq.pe122gs}
\end{equation}
where $N$ is given in eq.~(\ref{eq4.2.20a}).
\begin{equation}	
\Delta E=\Delta KE +\Delta PE=\Delta KE +\Delta PE_{12}+2\Delta PE_{13},\label{eq52}
\end{equation}
\begin{equation}
\Delta PE_{13}=\sum_{i=1}^4\Delta PE_{13i},\label{eq3.80}
\end{equation}
\begin{equation}
\Delta PE_{12}=\sum_{i=1}^3\Delta PE_{12i}.\label{eq3.81}
\end{equation}
These expressions for $\Delta E= \Delta KE +\Delta PE$, evaluated using the parameters given 
 in Table 1, are listed in Table 2.
\par
Analogous expressions for the first two excited states (relevant for $Z_n > 1$, \ie He-like systems) 
are given in ref. \cite{BarhamThesis}.

\end{document}